\documentclass[10pt,letterpaper,twocolumn]{article} %% two column, final layout

\usepackage{ol2}
\usepackage[draft]{hyperref}
\usepackage{amsmath}

\begin{document}

\twocolumn[ %% activate for two-column option

\title{Octave-spanning frequency comb generation in a silicon nitride chip}

%% For REVTeX it is possible to automate superscript and e-mail callouts with the superscriptaddress option; see REVTeX4 documentation.

\author{Yoshitomo Okawachi,$^{1,*}$ Kasturi Saha,$^1$ Jacob S. Levy,$^2$ Y. Henry Wen,$^1$ \\Michal Lipson,$^{2,3}$ and Alexander L. Gaeta$^{1}$}

\address{
$^1$School of Applied and Engineering Physics, Cornell University, Ithaca, New York 14853, USA \\ 
$^2$School of Electrical and Computer Engineering, Cornell University, Ithaca, New York 14853, USA \\
$^3$Kavli Institute at Cornell for Nanoscale Science, Cornell University, Ithaca, New York 14853, USA \\
$^*$Corresponding author: yo22@cornell.edu
}

\begin{abstract}We demonstrate a frequency comb spanning an octave via the parametric process of cascaded four-wave mixing in a monolithic, high-\textit{Q} silicon nitride microring resonator. The comb is generated from a single-frequency pump laser at 1562 nm and spans 128 THz with a spacing of 226 GHz, which can be tuned slightly with the pump power. In addition, we investigate the RF-noise characteristics of the parametric comb and find that the comb can operate in a low-noise state with a  30-dB reduction in noise as the pump frequency is tuned into the cavity resonance.\\  \end{abstract}

 ] %% activate for two-column option

\noindent The generation of broadband optical frequency combs is of great interest for numerous applications including spectroscopy, precision frequency metrology and optical clocks  \cite{Udem}. Many of the previous comb generation techniques rely on supercontinuum generation spanning 100's of THz in optical fibers using high-power mode-locked lasers \cite{Diddams}. The generation of comb spectra spanning an octave is highly desirable for full stabilization of the frequency comb through the well-established self-referencing technique \cite{Cundiff}. Recently, high \textit{Q}-factor optical microcavities have shown enormous potential as a platform for efficient nonlinear optical processes. One such process is four-wave mixing (FWM), which relies on a $\chi$$^{(3)}$ nonlinear optical process, where two pump photons annihilate to produce a symmetric signal-idler photon pair with respect to the pump frequency. In a cavity geometry, the FWM gain can lead to optical parametric oscillation, enabling the generation of multiple new wavelengths. The nonlinear enhancement enabled by such a high-\textit{Q} cavity allows for the possibility of broadband parametric frequency comb generation from a single-frequency cw laser \cite{DiddamsReview}. Parametric oscillation was first observed in silica microtoroids \cite{DelHayeNature, DelHayePRL, DelHayeArxiv}, and has since then has been demonstrated in CaF$_2$ and MgF$_2$ resonators \cite{Grudinin, Herr, Liang}, silica microspheres \cite{Agha}, silica-fiber Fabry-Perot cavities \cite{Braje}, high-index silica-glass microrings \cite{Razzari}, and silicon nitride microrings \cite{Levy, Foster, Okawachi, Ferdous}. Recently, stabilization of a 20-nm-broad comb \cite{DelHayePRL} and generation of an octave-spanning comb \cite{DelHayeArxiv} have been shown in silica microtoroids.

\begin{figure}[b]
\centerline{\includegraphics[width=7.5cm]{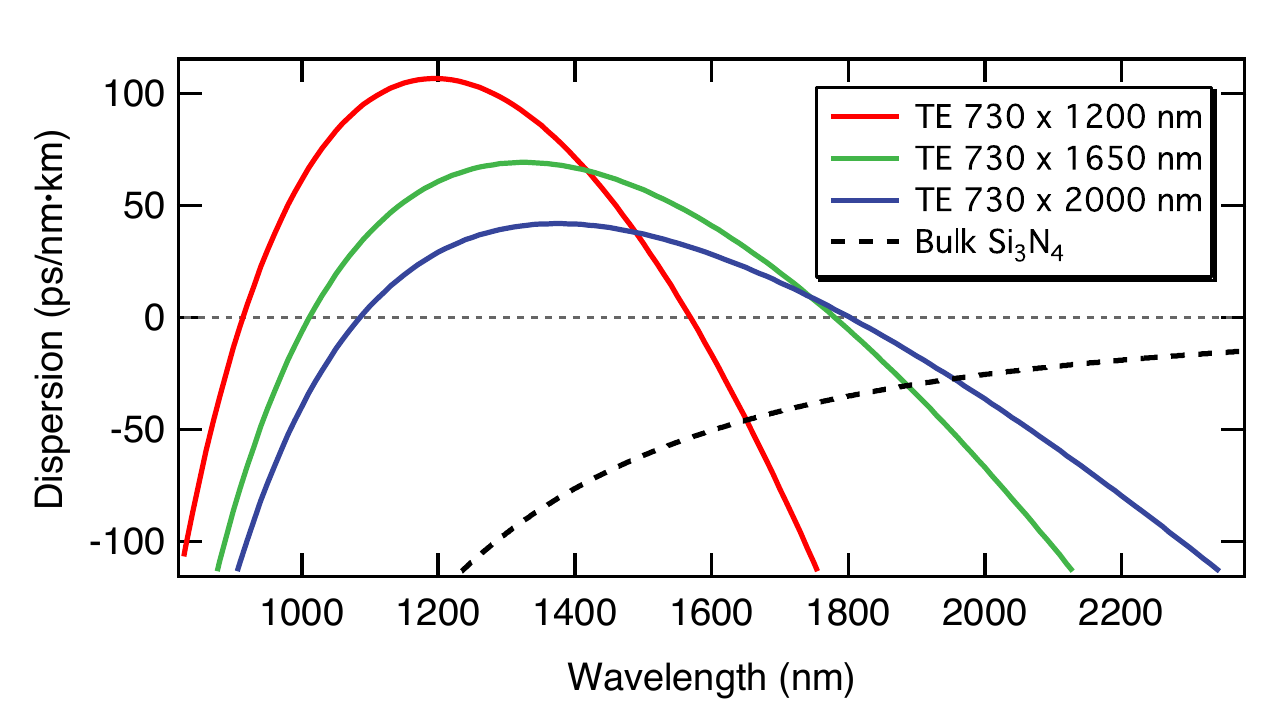}}
\caption{Dispersion simulations for the fundamental TE mode of a silicon-nitride waveguide with the height fixed at 730 nm and widths of 1200, 1650, and 2000 nm. The dashed curve indicates the dispersion for bulk silicon nitride.}
\label{fig:Disp}
\end{figure}

In this paper, we report the first demonstration of an octave-spanning parametric frequency comb in a silicon-based microring resonator. This silicon nitride platform is complementary-metal-oxide-semiconductor (CMOS) process compatible, and the devices are monolithic and sealed, such that pump coupling to the resonator and system operation are insensitive to the environment. Additionally, high-\textit{Q} silicon nitride resonators can be fabricated with  \textit{Q}-factors as high as 3$\times$10$^6$ \cite{Gondarenko}. The resonator is pumped at 1562 nm using a single-frequency pump laser, and we generate a parametric comb spanning from 1170 nm to 2350 nm, corresponding to a 128-THz span. This demonstration represents a significant step towards a stabilized, integrated frequency comb source that is robust and can be scaled to other wavelengths. 

\begin{figure*}[tb]
\centerline{\includegraphics[width=17.0cm]{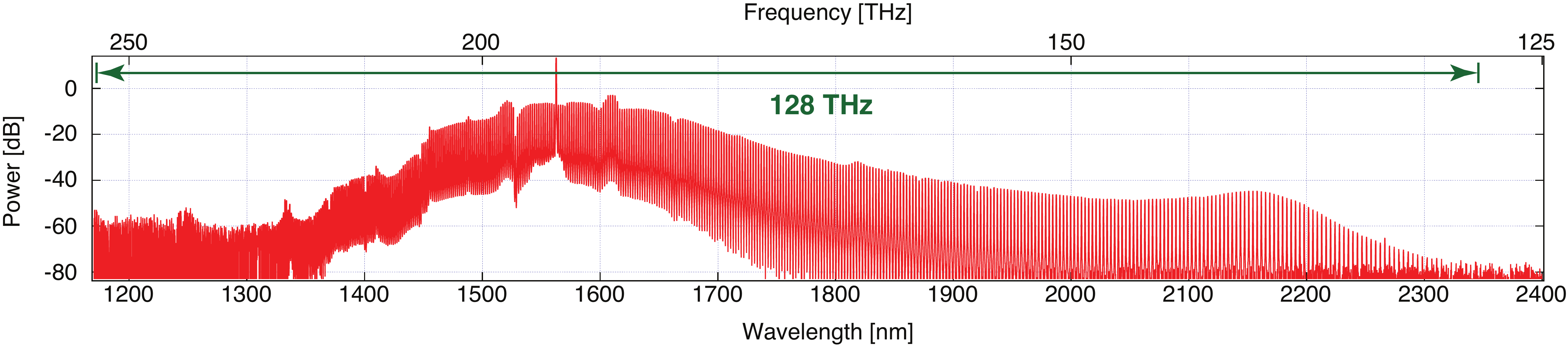}}
\caption{Optical spectrum of octave-spanning parametric frequency comb generated in a silicon nitride ring resonator.}
\label{fig:Octave}
\end{figure*}

FWM and parametric oscillation are enabled by the interplay between the nonlinearity and the group-Ávelocity dispersion (GVD) in the ring resonator. To achieve efficient phase matching, the pump laser is operated in the anomalous-GVD regime to compensate for the nonlinear phase mismatch. The cavity geometry results in an additional restriction, such that the pump, signal, and idler must each overlap with a cavity resonance. The signal/idler pair that oscillate at threshold are those that satisfy the energy conservation and phase matching conditions required for FWM \cite{Agha}. As the pump power inside the cavity is further increased, the signal/idler power grows, leading to cascaded FWM and multiple wavelength oscillations.   

The silicon nitride ring resonators are fabricated on a silicon wafer, and a 4-$\mu$m thermal oxide layer is initially grown for the undercladding. Next, a nitride film is deposited, and both the microring and the coupling waveguide are patterned and etched  monolithically in the silicon nitride layer. Finally, an oxide layer is deposited to form the top cladding \cite{Levy}. The relatively high index contrast between the nitride core and the oxide cladding allows for dispersion engineering through modification of the waveguide cross-section \cite{Turner, Tan}, and, since the dispersion is dependent on the cross-sectional size and shape, such engineering can be accomplished without changing resonator parameters such as the cavity free spectral range (FSR), enabling parametric comb generation at a wide range of pump wavelengths \cite{Okawachi}. To understand the waveguide conditions necessary to achieve broadband parametric comb generation, we theoretically calculate the GVD using a finite-element mode solver. Figure \ref{fig:Disp} shows the simulated dispersion for silicon nitride waveguides with a height of 730 nm, and widths of 1200, 1650, and 2000 nm. The curves indicate that large anomalous-GVD bandwidths spanning nearly an octave are possible with appropriate cross-section engineering.  

In our experiment, we amplify a single-frequency diode laser centered at 1562 nm with a erbium-doped fiber amplifier (EDFA) and inject it into a nanowaveguide using a lensed fiber. The input polarization is adjusted to quasi-TE using a fiber polarization controller. The nanowaveguide acts as the coupling waveguide for the microring resonator. Both the coupling waveguide and the microring have cross-sections of 725 nm by 1650 nm, and the microring has a 200-$\mu$m diameter. The power inside the coupling waveguide is 400 mW when the pump wavelength is detuned from a cavity resonance. The output is collected using an aspheric lens and sent to an optical spectrum analyzer (OSA).

\begin{figure}[b]
\centerline{\includegraphics[width=7.5cm]{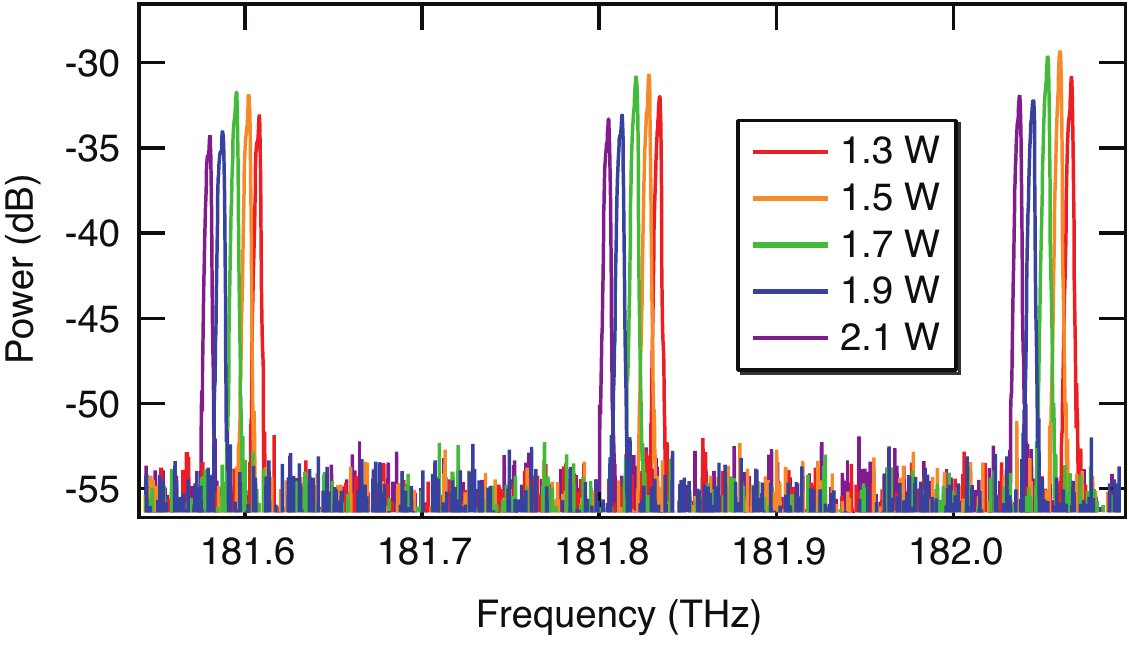}}
\caption{(Color online) Demonstration of tuning performance of the parametric comb. A 5-nm segment of parametric comb is monitored at various incident powers from 1.3 to 2.1 W. The frequency tuning range for this range of powers is 29 GHz, and we estimate the change in the comb spacing to be 36 MHz..}
\label{fig:tuning}
\end{figure}

The pump wavelength is tuned into a cavity resonance, such that a stable ``thermal lock" is achieved \cite{Carmon}. As the power in the microring is increased above threshold, the oscillation of a signal/idler pair occurs. Further increases in power lead to cascaded FWM, resulting in the generation of comb lines, and higher-order degenerate and nondegenerate FWM processes filling in adjacent cavity modes and increasing the comb line density. The parametric frequency comb spectrum from the microring resonator is shown in Fig. \ref{fig:Octave}. The total spectrum is acquired using two OSA's, with spectral ranges of 600-1700 nm and 1200-2400 nm. The comb lines are generated over a wavelength span of 1180 nm, from 1170 nm to 2350 nm, which corresponds to an octave of optical bandwidth.

We investigate the tuning performance of the parametric comb by monitoring a  5-nm section of the comb centered at 1650 nm using an optical spectrum analyzer. Figure  \ref{fig:tuning} shows that the three comb lines are red-shifted as the incident power is varied from 1.3 W to 2.1 W. As the pump power is increased, the resonance experiences a red-shift, which is primarily due to thermal effects. We observe a frequency tuning range of 29 GHz and estimate  that the comb spacing undergoes a 36-MHz shift over the range of incident powers used. This tuning range can be further increased with higher pump powers and higher \textit{Q}-factor resonators.  

\begin{figure}[t]
\centerline{\includegraphics[width=8.0cm]{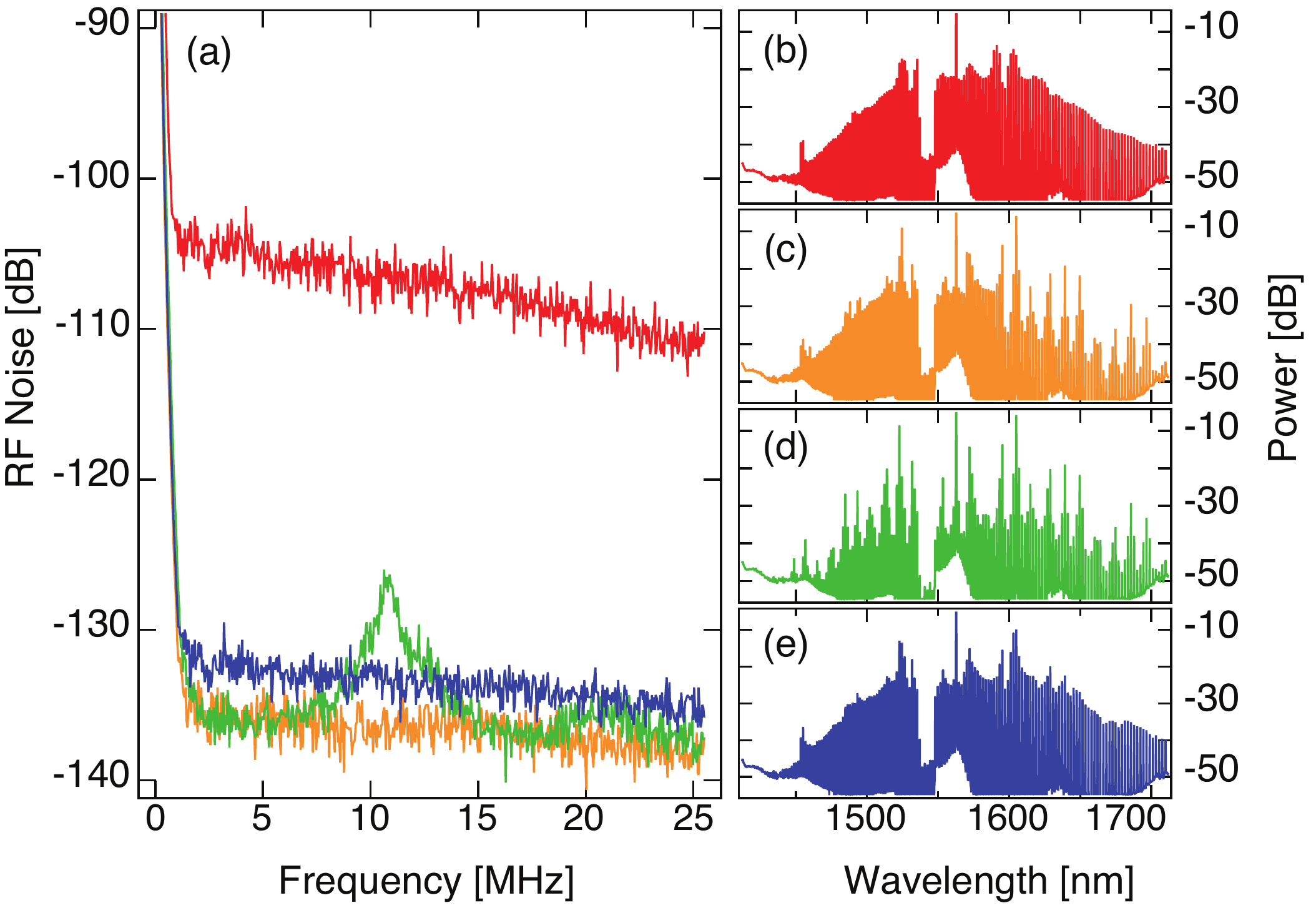}}
\caption{(Color online) (a) RF noise spectral of parametric frequency comb. A 9-nm portion of the optical spectrum is filtered from the comb for RF measurement. The noise is measured at four different pump detunings over 10 GHz as the pump is tuned into the cavity resonance. (b)-(e) show the corresponding optical spectra after the 9-nm section is filtered.}
\label{fig:RF}
\end{figure}

Lastly, we characterize the RF noise of the parametric comb. We use a wavelength division multiplexer to filter a 9-nm section of the comb spectrum that does not include the pump wavelength. The filtered portion is detected by a fast photodiode, and the electrical signal is sent to an RF spectrum analyzer. We simultaneously monitor the remaining spectrum using an optical spectrum analyzer, and Fig. \ref{fig:RF} shows the RF and optical spectra. As the pump wavelength is red-shifted into resonance, the RF noise decreases by 30 dB (red curve to yellow curve), and we observe modulations in the optical spectrum [Fig. \ref{fig:RF}(c) and (d)]. We observe a peak near 11 MHz, which we attribute to relaxation oscillations of the pump laser and we are currently investigating this issue further. As the pump is further red-shifted into the resonance, the the power in adjacent comb lines equalizes [Fig. \ref{fig:RF}(e)], while maintaining the low-noise state. At this time, we have not been able to determine the exact cause of this drop in noise although he hypothesize that it is an indication of the comb transitioning into a phase-locked state.

In conclusion, we demonstrate octave-spanning parametric frequency comb generation in a monolithically integrated silicon nitride microring resonator from a single-frequency cw laser. We achieve a 128-THz comb bandwidth with a comb spacing of 226 GHz. The broadband comb generation on-chip offers potential for a fully integrated, stabilized, compact optical frequency comb source with applications in spectroscopy, metrology, high-speed communications, and on-chip optical clocks. While the combs generated here were in the near IR, dispersion engineering of the silicon nitride ring resonator should enable generation of frequency combs at visible and at mid-IR wavelengths, allowing for unmatched flexibility in selecting the operating pump wavelength.

We acknowledge support from DARPA under the POPS and QuASAR programs, and by the Center for Nanoscale Systems, supported by the NSF and NYSTAR. This work was performed in part at the Cornell NanoScale Facility, a member of the National Nanotechnology Infrastructure Network, which is supported by the NSF (Grant ECS-0335765). We also acknowledge useful discussions with Mark Foster, Onur Kuzucu, and Ryan Lau. Y. O. and K. S. contributed equally to this work.

%\pagebreak


\begin{thebibliography}{99}

\bibitem{Udem} Th. Udem, R. Holzwarth, T. W. H\"{a}nsch, Nature \textbf{416,} 233 (2002).
\bibitem{Diddams} S. A. Diddams, D. J. Jones, J. Ye, S. T. Cundiff, J. L. Hall, J. K. Ranka, R. S. Windeler, R. Holzwarth, Th. Udem, and T. W. H\"{a}nsch, Phys. Rev. Lett. \textbf{84,} 5102 (2000).
\bibitem{Cundiff} S. T. Cundiff and J. Ye, Rev. Mod. Phys. \textbf{75,} 325 (2003).
\bibitem{DiddamsReview} T. J. Kippenberg, R. Holzwarth, and S. A. Diddams, Science \textbf{339}, 555 (2011).
\bibitem{DelHayeNature} P. Del'Haye, A. Schliesser, O. Arcizet, T. Wilken, R. Holzwarth, and T. J. Kippenberg, Nature \textbf{450,} 1214 (2007).
\bibitem{DelHayePRL} P. Del'Haye, O. Arcizet, A. Schliesser, R. Holzwarth, and T.J. Kippenberg, Phys. Rev. Lett. \textbf{101,} 053903 (2008).
\bibitem{DelHayeArxiv} P. Del'Haye, T. Herr, E. Gavartin, R. Holzwarth, and T. J. Kippenberg, arXiv:0912.4890v1 (2009).
\bibitem{Grudinin} I. S. Grudinin, N. Yu, and L. Maleki, Opt. Lett. \textbf{45,} 878 (2009).
\bibitem{Herr} T. Herr, C. Wang, P. Del'Haye, A. Schliesser, K. Hartinger, R. Holzwarth, and T. Kippenberg, in Quantum Electronics and Laser Science Conference, OSA Technical Digest (CD) (Optical Society of America, 2011), paper QTuF1. 
\bibitem{Liang} W. Liang, A. A. Savchenkov, A. B. Matsko, V. S. Ilchenko, D. Seidel, and L. Maleki, Opt. Lett. \textbf{36,} 2290 (2011).
\bibitem{Agha} I. H. Agha, Y. Okawachi, and A. L. Gaeta, Opt. Express \textbf{17,} 16209 (2009).
\bibitem{Braje} D. Braje, L. Hollberg, and S. Diddams, Phys. Rev. Lett. \textbf{102,} 193902 (2009). 
\bibitem{Razzari} L. Razzari, D. Duchesne, M. Ferrera, R. Morandotti, S. Chu, B. E. Little, and D. J. Moss, Nature Photon. \textbf{4,} 41 (2010).
\bibitem{Levy} J. S. Levy, A. Gondarenko, M. A. Foster, A. C. Turner-Foster, A. L. Gaeta, and M. Lipson, Nature Photon. \textbf{4,} 37 (2010).
\bibitem{Foster} M. A. Foster, J. S. Levy, O. Kuzucu, K. Saha, M. Lipson, and A. L. Gaeta, Opt. Express \textbf{19,} 14233 (2011).
\bibitem{Okawachi} Y. Okawachi, K. Saha, J. S. Levy, M. A. Foster, M. Lipson, and A. L. Gaeta, in CLEO:2011 - Laser Applications to Photonic Applications, OSA Technical Digest (CD) (Optical Society of America, 2011), paper CFK2.
\bibitem{Ferdous} F. Ferdous, H. Miao, D. E. Leaird, K. Srinivasan, J. Wang, L. Chen, L. T. Varghese, and, A. M. Weiner, arXiv:1103.2330 (2011).  
\bibitem{Gondarenko} A. Gondarenko, J. S. Levy, and M. Lipson, Opt. Express \textbf{17,} 11366 (2009). 
\bibitem{Turner} A. C. Turner, C. Manolatou, B. S. Schmidt, M. Lipson, M. A. Foster, J. E. Sharping, and A. L. Gaeta, Opt. Express \textbf{14,} 4357 (2006).
\bibitem{Tan} D. T. H. Tan, K. Ikeda, P. C. Sun, and Y. Fainman, Appl. Phys. Lett. \textbf{96,} 061101 (2010).
\bibitem{Carmon} T. Carmon, L. Yang, and K. J. Vahala, Opt. Express \textbf{12,} 4742 (2004).

\end{thebibliography}
\end{document}